\documentstyle[times,graphicx]{mn}
\newcommand{\fig}[1]{\includegraphics[width=\hsize]{#1}}
\title[Efficient Multi-Gaussian Expansion of galaxies]{Efficient Multi-Gaussian Expansion of galaxies$^\ddagger$}
\author[M. Cappellari]{Michele Cappellari
\thanks{E-mail: cappellari@strw.leidenuniv.nl}
\thanks{European Space Agency external fellow.}\\
    Leiden Observatory, Postbus 9513, 2300 RA Leiden, The Netherlands}

\pagerange{\pageref{firstpage}--\pageref{lastpage}}
\pubyear{2002}

\begin{document}

\label{firstpage}

\maketitle

\begin{abstract}
We describe a simple, efficient, robust and fully automatic algorithm for the determination of a Multi-Gaussian Expansion (MGE) fit to galaxy images, to be used as a parametrization for the galaxy stellar surface brightness. In most cases the least-squares solution found by this method essentially corresponds to the \emph{minimax}, constant relative error, MGE approximation of the galaxy surface brightness, with the chosen number of Gaussians. The algorithm is well suited to be used with multiple resolution images (e.g., \emph{Hubble Space Telescope} [HST] and ground-based). It works orders of magnitude faster and is more accurate than currently available methods. An alternative, more computing intensive, fully linear algorithm, that is \emph{guaranteed} to converge to the smallest $\chi^2$ solution, is also discussed. Examples of MGE fits are presented for objects with HST or ground-based photometry, including galaxies with significant isophote twist.
\end{abstract}


\section{Introduction}
\footnotetext[3]{Based on observations made with the NASA/ESA {\it Hubble Space Telescope}, obtained from the Data Archive at the Space Telescope Science Institute, which is operated by the AURA, Inc., under NASA contract NAS 5-26555.}

\setcounter{footnote}{4}

Digital images of galaxies which are small by today's standards, may still contain over one million pixels. It is natural to try to distill the information contained in this large number of pixels into a set of quantities that can be used easily and efficiently  to derive information on the physical properties of the observed object. The most common approach currently consists in fitting ellipses of increasing semimajor axis to the galaxy images (e.g., Carter 1978; Kent 1984; Lauer 1985; Jedrzejewski 1987; Franx, Illingworth \& Heckman 1989; Peletier et al.\ 1990). Intensity profiles, ellipses shape and deviation from ellipses, parametrized by a few Fourier terms, are obtained with these methods. A nice feature of this parametrization is that it gives a physically meaningful description of the galaxy photometry in terms of the ellipticity, diskyness and boxyness at each radius.

One limitation of the ellipse fitting methods however is that strong deviations of the isophotes from ellipses cannot be easily modeled with a small number of Fourier terms. This makes these methods not well suited to the photometric modeling of multicomponent objects such as lenticulars and spirals, which have a bulge, a main disk and often an embedded nuclear disk or a bar. More importantly, if the deviations from ellipses are non negligible it becomes non trivial to use the fitted isophotes for the construction of realistic galaxy dynamical models due to the complexity of deprojection and the evaluation of the gravitational potential.

In this paper we focus on a completely different approach to the photometric modeling, the Multi-Gaussian Expansion (MGE) method, that overcomes both of these limitations.
The original idea of the application of the MGE method to the problem of the deconvolution and deprojection of galaxy images comes from Bendinelli (1991). This idea was generalized to the non-spherical case and made applicable to real galaxies by Monnet, Bacon \& Emsellem (1992) and further developed by Emsellem, Monnet \& Bacon (1994a) and Emsellem, Dejonge \& Bacon (1999).

The MGE method consists of a series expansion of galaxy images using two-dimensional (2D) Gaussian functions. The Gaussians have the beneficial properties that both the convolution (e.g., to take seeing or PSF effects into account) and the deprojection (to derive  the intrinsic stellar luminosity density from the observed galaxy photometry) can be performed analytically in a simple and efficient manner.
As shown by Emsellem et al.\ (1994a), many other dynamical and photometric quantities can be evaluated easily and accurately when the density is expressed in MGE form. For example, the MGE potential can be computed with a single integration, as opposed to the two that are required when the intrinsic density is stratified on similar triaxial ellipsoids, and three in the general case. In the case of $f(E,L_z)$ axisymmetric MGE dynamical models the velocity moments predicted from the Jeans equations, already projected onto the sky, can be expressed with only a double integration. The even part of the distribution function can also be easily retrieved from an MGE density distribution via the Hunter \& Qian (1993) formalism.

The MGE parametrization is one of the few simple parame\-trizations that are general enough to reproduce the surface brightness of realistic multicomponent objects (e.g., spirals with multiple disks). It has already been used for the modeling of a number of galaxies (Emsellem et al.\ 1994b, 1996, 1999; Emsellem 1995; van den Bosch et al.\ 1998; van den Bosch \& Emsellem 1998; Cretton \& van den Bosch 1999).

What is however still missing from the MGE machinery is an accurate, easy to use, generally available, robust and automatic algorithm for the determination of an MGE expansion from galaxy images. In this paper we present such an algorithm. The method we describe here brings the MGE fitting phase of a dynamical modeling process to the same level of the LOSVD extraction phase (e.g., van der Marel \& Franx 1993) in terms of robustness and ease of use.

This paper is organized as follows. In Section~2 we define our notation and give some useful formulae for the application of MGE models to the dynamical modeling of galaxies. In Section~3 we discuss our new MGE fitting method. In Section~4 we give information on the availability of the software implementing the methods discussed in this paper. Some conclusions are given in Section~5.

\section{MGE formalism}

In this section we rederive, in a slightly different notation, results from Monnet et al.\ (1992) and Emsellem et al.\ (1994a) that are usually needed for the construction of dynamical models, starting from the MGE photometric models we describe in this paper. The case of Gaussians with the same centre and circular PSF is considered here, while the general case is discussed in detail in Emsellem et al.\ (1994a).

Let $(x',y',z')$ be a system of coordinates centred on the galaxy nucleus, with the $z'$ axis pointing towards the observer. The MGE projected surface brightness can be written as
\begin{equation}
\Sigma(R',\theta') = \sum_{j=1}^N
{\frac{L_j}{2\pi\sigma_j^2 q_j'} \exp
\left[
    -\frac{1}{2\sigma_j^2}
    \left(x'^2_j + \frac{y'^2_j}{q'^2_j} \right)
\right]},
\label{eq:surf_twist}
\end{equation}
with
\begin{equation}
\left\{ \begin{array}{l}
 x'_j  = R' \sin(\theta'-\psi_j),  \\
 y'_j  = R' \cos(\theta'-\psi_j),  \\
 \end{array} \right.
\end{equation}
where $(R',\theta')$ are the polar coordinates on the plane of the sky $(x',y')$. Here $N$ is the number of the adopted Gaussian components, having total luminosity $L_j$, observed axial ratio $0\le q'_j\le1$, dispersion $\sigma_j$ along the major axis, and position angle (PA) $\psi_j$, measured counterclockwise from the $y'$ axis to the major axis of the Gaussian.

Given that the convolution, deprojection and potential calculations, can be carried out separately for each single Gaussian component of equation~(\ref{eq:surf_twist}), in what follows the index of the Gaussian is not written explicitly and only one Gaussian is considered. The results for the complete MGE model are obtained by summing the contributions over the $N$ Gaussians components.

\subsection{PSF convolution}

The MGE surface brightness has to be convolved with the instrumental PSF before comparison with the observed surface brightness.
Assume the normalized PSF can be written as the sum of $M$ circular Gaussians
\begin{equation}
{\rm PSF}(R') = \sum_{k=1}^M
{\frac{G_k}{2\pi\sigma_k^{\star 2}} \exp
\left(
    -\frac{R'^2}{2\sigma_k^{\star 2}}
\right)},
\end{equation}
with $\sum_{k=1}^M {G_k}=1$. The total luminosity $L$ of one Gaussian, having surface brightness $\Sigma$, does not change after convolution $\overline{\Sigma}=\Sigma\otimes{\rm PSF}$ with the PSF. The convolved Gaussian can then be written explicitly as
\begin{equation}
\overline{\Sigma}(R',\theta') = L \sum_{k=1}^M
{\frac{G_k}{2\pi\overline\sigma_k^2\, \overline q' _k}  \exp
\left[
    -\frac{1}{2\overline\sigma_k^2}
    \left(x'^2 + \frac{y'^2}{\overline q'^2 _k} \right)
\right]},
\label{eq:psf1}
\end{equation}
where
\begin{equation}
\left\{ \begin{array}{l}
 \overline\sigma^2_k  = \sigma^2  + \sigma_k^{\star 2}  \\
 \overline\sigma^2_k\, \overline q'^2_k =  \sigma^2 q'^2  + \sigma_k^{\star 2}  \\
 \end{array} \right.
\label{eq:psf2}
\end{equation}

\subsection{Deprojection}

The deprojection of a galaxy surface brightness distribution is generally non unique (e.g., Rybicki 1986; Franx 1988). The MGE deprojection discussed here represents a possible solution to the problem. It generally produces smooth and natural-looking intrinsic densities, consistent with the observed photometry. The nonuniquness of the triaxial deprojection on a set of real galaxies and its effects on the dynamical modeling will be the subject of a coming paper (Cappellari \& de Zeeuw, in preparation).

\subsubsection{Triaxial case}

If a good MGE fit to the photometry can only be obtained by allowing the PA $\psi_j$ of the Gaussians to be different from each other, then the object can \emph{not} be axisymmetric. The following deprojection method can be used to produce an intrinsic density model consistent with the observations.

Consider one single Gaussian component taken from equation~(\ref{eq:surf_twist}). Since the isophotes of this component are similar ellipses it can be deprojected by assuming the intrinsic density is stratified on triaxial ellipsoids (Contopoulos 1956; Stark 1977; Binney 1985). The intrinsic luminosity density of this component can be written again as a Gaussian with the same total luminosity $L$
\begin{equation}
\rho(x,y,z) =
\frac{L}{(\sqrt{2\pi}\, \sigma)^3 p q} \exp
\left[
    -\frac{1}{2\sigma^2}
    \left(x^2+\frac{y^2}{p^2}+\frac{z^2}{q^2} \right)
\right].
\label{eq:density}
\end{equation}
Here $(x,y,z)$ is a system of coordinates as in Binney (1985), centred on the Gaussian centre and aligned with its principal axes. The axes are assumed to be such that the longest axis of the Gaussian lies along the $x$-axis and the shortest axis along the $z$-axis, so that $0\le q\le p\le 1$. The two usual polar coordinates $(\theta,\phi)$ define the orientation of the line-of-sight with respect to the principal axes of the object. Another angle $\psi$ is required to specify the rotation of the object around the line-of-sight, and uniquely define the orientation in space of the $(x,y,z)$ coordinate system. This angle $\psi$ is defined by the requirement that the $z$ axis project onto the $y'$ axis, or equivalently that the $x'$ axis lies in the $(x,y)$ plane.

Once the viewing angles $(\theta,\phi,\psi)$ have been \emph{assumed}, given the observed axial ratio $q'$ for the Gaussian, the intrinsic axial ratios $p$ and $q$ can be found using relations valid for general ellipsoidal bodies (e.g., de Zeeuw \& Franx 1989)
\begin{eqnarray}
1\! -\! q^2\! =\! \frac{\delta'
     \left[ 2\cos 2\psi\! +\!
       \sin 2\psi (  \sec \theta \cot \phi\! -\! \cos \theta \tan \phi)  \right] }{2\sin^2 \theta
     \left[ \delta' \cos \psi
        ( \cos \psi\!  +\!
          \cot \phi \sec \theta \sin \psi  )\! -\! 1  \right] },\\
p^2\! -\! q^2\! =\!
  \frac{ \delta'
     \left[ 2\cos 2\psi\!  +\!
       \sin 2\psi ( \cos \theta \cot \phi\! -\! \sec \theta \tan \phi )  \right] }{2\sin^2 \theta
     \left[ \delta' \cos \psi
        ( \cos \psi\!  +\!
          \cot \phi \sec \theta \sin \psi  )\! -\! 1 \right] },
\end{eqnarray}
where $\delta'=1-q'^2$.
It should be noted that in general a solution may not be found for all assumed viewing angles (e.g., Bertola et al.\ 1991). If a triaxial deprojection is sought for the galaxy, then the viewing angles have to be the same for all the Gaussians and the permissible viewing angles for the object, assumed to be triaxial, will lie in the intersection of the permissible viewing angles of each Gaussian (cf. Williams 1981).

\subsubsection{Axisymmetric case}

If an acceptable MGE model can be obtained by fixing the PA $\psi_j$ of all the Gaussians to the same value $\psi$, then the object can be deprojected by assuming it is axisymmetric (triaxial solutions however can \emph{not} be excluded [cf.\ Franx 1988]). In this case the formulae of the previous section simplify considerably.

In the oblate axisymmetric case ($p=1$), once an inclination $i>0$ (for $i=0$ the deprojection is degenerate) has been assumed for the galaxy ($i=90^\circ$ corresponding to edge-on), one has:
\begin{equation}
    q^2=\frac{q'^2-\cos^2 i}{\sin^2 i},
\end{equation}
while in the prolate axisymmetric case ($p=q$)
\begin{equation}
    q^2=\frac{\sin^2 i}{1/q'^2-\cos^2 i}.
\end{equation}

The oblate MGE deprojection above is only defined if $\cos^2 i<q'^2_j$ for all Gaussians. This means that the flattest Gaussian in an MGE fit dictates the minimum possible inclination for which the MGE model can be used in a dynamical model. However since the Gaussian components don't necessarily have physical significance one don't want to base any conclusion of a dynamical model on this minimum inclination.

One simple way to avoid this problem consists in trying to increase the minimum axial ratio $q'$ allowed in the MGE fit until the $\chi^2$ increases significantly or the galaxy contour cannot be accurately reproduced any more with any number of Gaussians. This $q'_{\rm max}$ will provide a good estimate of the minimum inclination for which the given galaxy, assumed to be axisymmetric, can be deprojected.

Independently from an actual MGE fit, the limit to the above $q'_{\rm max}$ is set by the axial ratio $q'$ of the ellipse having the same curvature of the galaxy isophotes along the major axis, at the point of maximum curvature. In fact there has to be at least one Gaussian flat enough to reproduce the maximum curvature along the major axis.

\subsection{Gravitational potential}

The potential generated by a triaxial Gaussian mass distribution of the form given in equation~(\ref{eq:density}), can be evaluated using the classical Chandrasekhar (1969) formulae for densities stratified on similar concentric ellipsoids (see Binney \& Tremaine 1987,  p.~61). In the MGE case the potential can be written as a one-dimensional (1D) integral
\begin{eqnarray}
    \lefteqn{\Phi(x,y,z) = \frac{\sqrt{\frac{2}{\pi}}\, G \Upsilon L}{\sigma}\ \times} \nonumber\\
 & &    \int_0^1
        {
        \frac{
        \exp \left[-\frac{T^2}{2\sigma^2} \left( x^2 + \frac{y^2}{1-\delta T^2} + \frac{z^2}{1-\epsilon T^2} \right) \right]
        }{
        \sqrt{(1-\delta T^2)(1-\epsilon T^2)}
        }
        } {\rm d}T,
\label{eq:pot_tri}
\end{eqnarray}
where $\delta=1-p^2$, $\epsilon=1-q^2$, G is the gravitational constant and $\Upsilon$ is the mass-to-light-ratio.
For oblate axisymmetric models ($p=1$) this reduces to
\begin{equation}
\Phi(R,z) = \frac{\sqrt{\frac{2}{\pi}}\, G \Upsilon L}{\sigma}\!
   \int_0^1
        {
        \frac{
        \exp \left[-\frac{T^2}{2\sigma^2}\! \left( R^2\! +\! \frac{z^2}{1-\epsilon T^2} \right) \right]
        }{
        \sqrt{1-\epsilon T^2}
        }
        } {\rm d}T.
\label{eq:pot_axis}
\end{equation}
In the spherical case ($p=q=1$) to
\begin{equation}
    \Phi(R) = \frac{G \Upsilon L}{R}\ {\rm erf}\!\left(\frac{R}{\sqrt{2}\ \sigma}\right).
\end{equation}
Already for $8\sigma\la R$ the error function ${\rm erf}[R/(\sqrt{2}\ \sigma)]=1$ to 16 digits accuracy and the spherical MGE potential can be replaced with high precision by the potential of a point mass enclosing the total mass ${\cal M}=\Upsilon L$ of the Gaussian component.

Since the $\sigma$ of the Gaussians in an MGE fit span various orders of magnitude in size, asymptotic expressions can also be conveniently used to estimate the triaxial or axisymmetric potential of equations~(\ref{eq:pot_tri}) and (\ref{eq:pot_axis}) at very small and very large radii as done in Cappellari et al.\ (2002).

\section{MGE fitting}

\subsection{The brute-force 2D approach}

We start by considering the MGE fit to galaxies without isophote twists.
The determination of an axisymmetric MGE parametrization from the photometry, namely the fit of a sum of Gaussians with the same centre and PA, to a galaxy image seems to be a very simple problem that can be efficiently solved with general nonlinear minimization algorithms.

We assume for simplicity that the PA $\psi$ of all Gaussians is known. We don't fit $\psi$  since it is very easily measured before fitting. We found it fast and accurate to determine it from the weighted second moments of the surface brightness above a given level, as done by automatic galaxy extraction packages (e.g., Bertin \& Arnouts 1996). It may also be measured with standard photometric programs, or may be dictated by other requirements in the dynamical model (in general a more accurate $\psi$  is required in the centre where the kinematical data are available). We will also always assume that the MGE models are analytically convolved with a circular MGE PSF before comparing with the data using Equations~(\ref{eq:psf1}) and (\ref{eq:psf2}).

The straightforward MGE fit approach consists of fitting the nonlinear function of equation~(\ref{eq:surf_twist}) to the pixels of the image by minimizing the $\chi^2$ (pixel integration can be taken into account in the very innermost few pixels)
\begin{equation}
\chi^2  = \sum_{i,j}
    \left[
        \frac{C_{i,j} - \Sigma(x'_i,y'_j)}
             {\Delta_{i,j}}
    \right]^2,
\label{eq:chi2}
\end{equation}
where $C_{i,j}$ are the pixel counts, $\Delta_{i,j}$ is a corresponding weight factor, depending on the photometric and systematic errors, and the sum is done over all (good) image pixels. Positivity constraints are generally enforced on the $L_j$, although in principle only the total surface density needs to be positive everywhere.

Although the convergence can be improved by starting the fit with only a few Gaussians and then adding new components as long as the residuals decrease appreciably (Emsellem et al.\ 1994a), we have discovered that this brute force approach does not work well in practice for two main reasons:
\begin{enumerate}
\renewcommand{\theenumi}{(\arabic{enumi})}
    \item the nonlinear function $\chi^2$ above has a large number of local stationary points. In fact suppose that we have found a minimum of the $\chi^2$ with $N$ Gaussians. A new configuration with $N+1$ Gaussians, where the new Gaussian is equal to one of those previously found, is a saddle point for the $\chi^2$. Consequently even state-of-the-art nonlinear fitting algorithms have a high chance of stopping or slowing down considerably in a narrow valley in the $3 N$-dimensional space (with $N\approx10-15$) close to one of these many stationary points, unless the initial guess is very near the true minimum;

    \item the ``correct'' weighting of the pixel measurements tends to produce a model that underestimates the nuclear density due to the fact that the total $\chi^2$ of equation~(\ref{eq:chi2}) is dominated by the contribution of many pixels at large radii. This is \emph{not} what is needed for an accurate dynamical model of the nucleus.
\end{enumerate}

The net result is that in the above approach the fit proceeds by trial and error, with a lot of manual intervention. An automatic fit generally takes many hours on a 1~GHz PC and usually converges to a solution that can still be improved manually.

\subsection{An efficient 1D algorithm\label{sec:1d}}

To solve the convergence problems of an MGE fit we start by analysing the peculiarities of a fit to a 1D profile, as can be obtained by measuring the photometry of a galaxy with constant ellipticity. Since galaxy profiles are generally well described by multiple power-law regimes (e.g., Lauer et al.\ 1995), in this section we start by trying to satisfy the natural requirement that the fit has to ``look good'' on a log-log plot. To obtain this result we do the following:
\begin{enumerate}
\renewcommand{\theenumi}{(\arabic{enumi})}
    \item we sample the profile logarithmically in radius. This is usually done in galaxy photometry to improve the S/N at large radii while still maintaining all the information at the smaller radial scales;

    \item we require the fit to optimise the \emph{relative} error, namely we minimize
    \begin{equation}
        \chi^2  = \sum_{j=1}^M \left[ \frac{C_j-\Sigma(R_j)}{C_j} \right]^2,
        \label{eq:chi2new}
    \end{equation}
    where $M$ is the number of photometric data points $C_j$ in the profile, $R_j$ is the corresponding isophote major axis and $\Sigma(R)$ is the 1D version of equation~(\ref{eq:surf_twist}).
\end{enumerate}

\begin{figure}
\fig{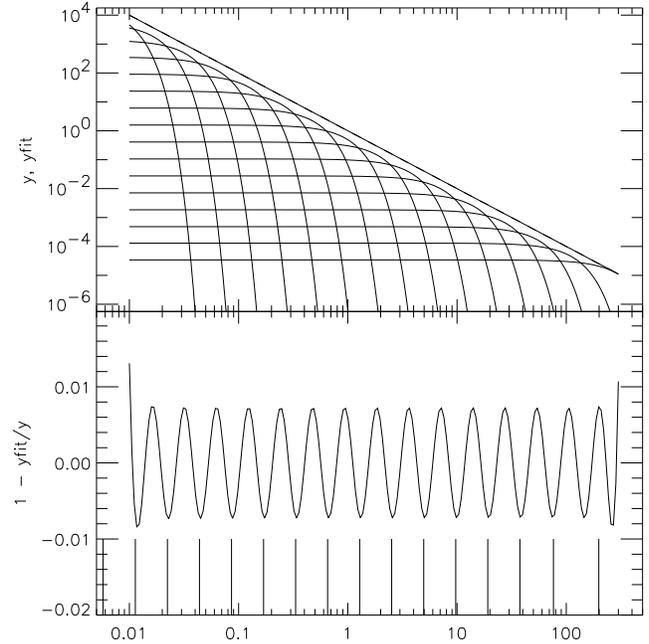}
\caption{1D MGE fit to the power-law profile $R^{-2}$. \emph{Top panel:} the profile and the $N=16$ best-fitting Gaussian components are shown. \emph{Bottom panel:} radial variation of the relative error in the profile fit. Note that the maximum error is reached precisely $2N+1$ times, with alternating signs. The vertical lines indicate the $\sigma_j$ of the Gaussians shown in the upper panel.\label{fig:mge_fit_1d_powerlaw}}
\end{figure}

The reason why Gaussians are so successful in reproducing power-law profiles lies in the fact that every Gaussian provides a very localized contribution to the profile, only for radii close to its $\sigma_j$. In fact 1) at small radii ($R\ll\sigma_j$) Gaussians become very rapidly constant, while power-laws continue to rise, 2) at large radii  ($R\gg\sigma_j$) Gaussians go to zero much more rapidly than a power-law. It appears clear that the MGE best-fitting solution for a logarithmically sampled power-law profile consists of a series of Gaussians that are also logarithmically spaced, both in $\sigma_j$ and in height, within the ranges imposed by the data (see Fig.~\ref{fig:mge_fit_1d_powerlaw}).

\begin{figure}
\fig{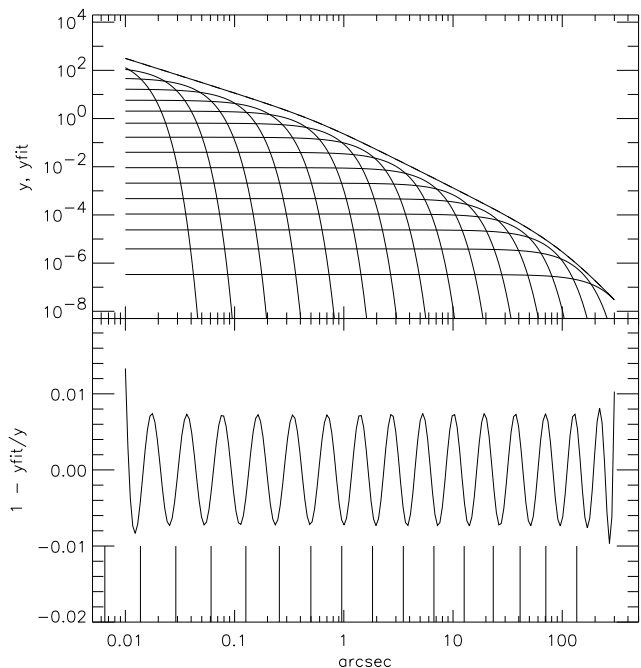}
\caption{Same as in Fig.~\ref{fig:mge_fit_1d_powerlaw} for the triple power-law profile  of the intrinsic density of M32, as determined by van der Marel et al.\ (1998). With $N=16$ Gaussians the maximum error is 0.7 per cent (excluding border effects) in the range 0\farcs01--300\arcsec, over ten orders of magnitude in the density.\label{fig:mge_fit_1d_m32}}
\end{figure}

\begin{figure}
\fig{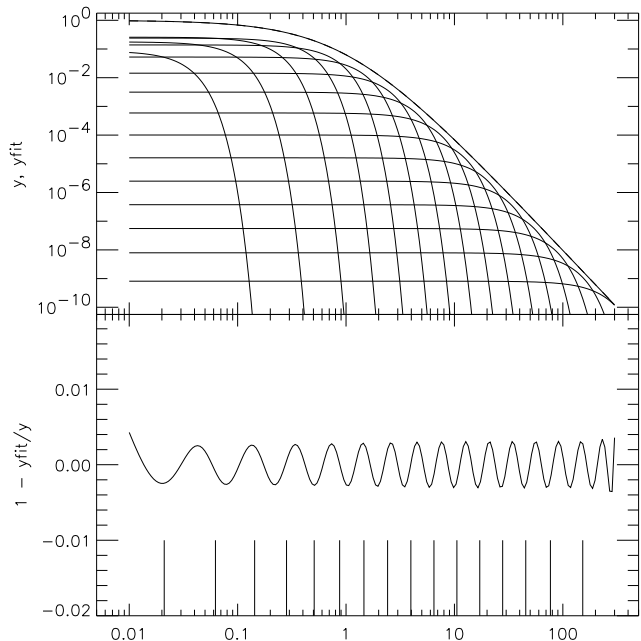}
\caption{Same as in Fig.~\ref{fig:mge_fit_1d_powerlaw} for the double power-law profile with flat core $(1 + R)^{-4}$. The maximum error is 0.3 per cent (excluding border effects) over ten orders of magnitude.
\label{fig:mge_fit_1d}}
\end{figure}

We now show that this solution is the true \emph{global} minimum.
In fact our least-squares MGE solution, with positive Gaussians, for a given decreasing function of radius, in a given interval, is characterized by the condition that the maximum relative error is reached precisely $2N+1$ times, with alternating signs (bottom panel of Fig.~\ref{fig:mge_fit_1d_powerlaw}). A similar condition also exists for the unique \emph{minimax}\footnote{The solution minimizing the maximum absolute error.} solution of other approximating functions e.g., Chebyshev polynomials and rational functions (see e.g., Press et al.\ 1992).
The fact that the minimax condition is satisfied for our MGE fits shows that our algorithm is able to converge to the minimax MGE approximation of a power-law.
The same minimax condition can be verified in the case of MGE fits to more realistic profiles composed of multiple power-law regimes (Fig.~\ref{fig:mge_fit_1d_m32}), and essentially for any function that appears smooth in a log-log plot. Here the Gaussians will not be precisely logarithmically spaced in the $\sigma_j$, but will be more closely spaced where the profile is steeper (Fig.~\ref{fig:mge_fit_1d}). Also in these cases however the least-squares solution essentially provides  the minimax MGE approximation of the given function.

We emphasize the fact that the logarithmic sampling of the profiles is a necessary condition for the least-squares solution to coincide, within the numerical approximations, with the minimax solution for the power-law. A linear sampling would have produced an MGE fit with smaller errors at large radii and larger errors towards the centre.

With real photometric data the above condition for the minimax solution cannot be always verified, due to the noise and to the fact that the observed profile is known at a finite number of points. The convergence to a global solution however can still be recognized, by simple visual inspection of the result, from the fact that every Gaussian has to give a significant contribution to a specific part of the profile: a Gaussian that remains always lower than any other Gaussian is generally an indication that the solution is not the global minimum (unless the profile becomes very flat).

To determine the fits of Fig.~\ref{fig:mge_fit_1d_powerlaw}--\ref{fig:mge_fit_1d} we made use of the important fact that equation~(\ref{eq:surf_twist}) is \emph{linear} in the Gaussian total luminosity $L_j$. Since a global solution for the linear variables $L_j$ can be found easily for any given combination of the nonlinear ones $\sigma_j$, it is crucial to separate the treatment of these two sets of variables. In practice our 1D MGE fitting algorithm consists of the following steps:
\begin{enumerate}
    \item we start with the $\sigma_j$ logarithmically spaced in radius between the maximum radius of the measured photometric data $R_{\rm max}$ and a minimum radius $R_{\rm min}=0.75\;\sigma_{\rm PSF}$. We know this is a good guess for general multiple power-law profiles;

    \item we solve the nonlinear least-squares minimization problem for the $\log \sigma_j$, imposing the constraints that $R_{\rm min}<\sigma_j<R_{\rm max}$. The actual computation is done using the robust Levenberg-Marquardt implementation due to Mor\'e, Garbow \& Hillstrom (1980)\footnote{We have used an IDL porting of the MINPACK-1 code by Craig B.\ Markwardt that is available from  http://cow.physics.wisc.edu/$\sim$craigm/idl}. The adopted implementation accepts bound constraints on the variables and is able to deal with degenerate Jacobians. The latter is a necessary feature since some of the variables $L_j$ may be zero in the final solution;

    \item for every choice of the $\sigma_j$ variables, during the above iterative optimisation process, we find the best-fitting value for the $L_j$, with the constraints $L_j\ge0$. This is a Non-Negative Least-Squares (NNLS) problem that can be efficiently solved using the Lawson \& Hanson (1974) algorithm\footnote{We have ported to IDL their 1995 Fortran~90 version of the Bounded-Variables Least-Squares (BVLS) code, a generalization of NNLS, available on http://www.netlib.org.}.
\end{enumerate}

In case a solution with negative Gaussians is needed (e.g., to reproduce a profile that is \emph{not} monotonically decreasing with radius) the solution of the linear system with NNLS, in step (iii) above, must be replaced by a solution using the Singular Value Decomposition (SVD) to provide a zero order (i.e.\ minimum norm) regularization on the solution (e.g., Press et al.\ 1992). Other linear regularization constraints could also be enforced on the solution. A direct non-regularized solution of the linear system is extremely ill-conditioned and will produce a result containing elements $L_j$ of very large absolute value and different sign. An MGE model obtained in this way presents the risk of cancellation of significant digits when deriving any physical quantity.

\begin{figure}
\fig{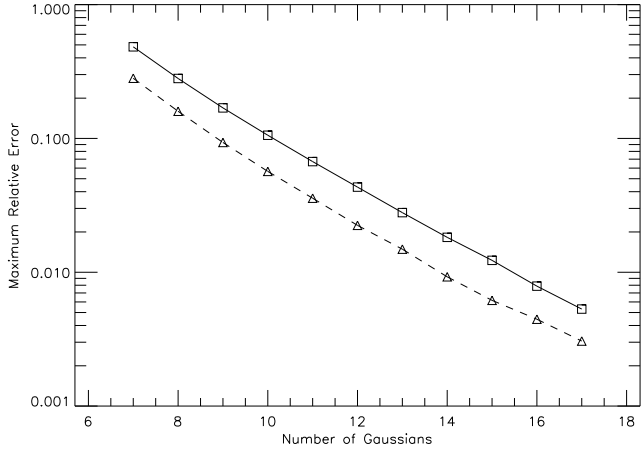}
\caption{Variation of the maximum relative error in the MGE fit of Fig.~\ref{fig:mge_fit_1d_m32} (solid line) and in the fit of Fig.~\ref{fig:mge_fit_1d} (dashed line) as a function of the number of Gaussians adopted in the expansion.
\label{fig:mge_error}}
\end{figure}

Fig.~\ref{fig:mge_error} shows that the maximum relative error in an MGE fit to multiple power-law profiles decreases almost exponentially with increasing number of Gaussians, approximately as fast as $\epsilon\propto e^{-0.4 N}$. This is not surprising, given that every Gaussian diverges exponentially from the power-law starting from the point of closest approach. Although the actual numbers will depend on the profiles and on the adopted radial ranges, these examples provide an estimate of the number of Gaussians that are required to reach a given accuracy in the fit, and explain why 10--15 Gaussians generally produce a very good MGE fit to realistic Galaxy profiles.

We finally mention that an asymptotic power-law nuclear cusp can also be imposed on the MGE profile as described in Emsellem et al.\ (1999). Although this choice breaks  the elegance of the MGE parametrization somewhat, this detail does not change in any way the essence of our fitting algorithm.

\subsection{Extension of the 1D algorithm to 2D images without isophote twist\label{sec:2d}}

We now extend the 1D algorithm described above to the fit of actual 2D images. We perform this extension by starting from the obvious requirement that the 2D algorithm provides \emph{exactly} the same results of the 1D case (to be precise the same $L_j$ and $\sigma_j$) when it is used to fit a 2D image that presents the same major axis profile of the 1D case and has perfectly elliptical isophotes with constant axial ratio $q'$ and PA.

\begin{figure}
\fig{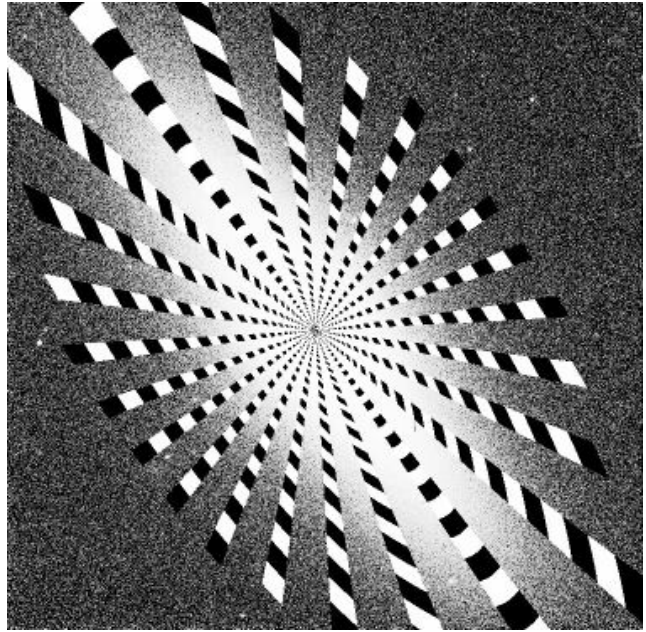}
\caption{This image illustrates the sectors that have been used to produce an MGE model for the WFPC2/F814W image of the flattened S0 galaxy NGC~4342. In the fit $N_{\rm sec}=19$ sectors have been used, spaced in angle by $5^\circ$ and covering the whole galaxy from the major to the minor axis. In this figure only one every three sectors is shown. These are the sectors from which the profiles shown in Fig.~\ref{fig:n4342_profiles} have been extracted. Note that the radial sampling along the sectors follows a characteristic isophote of the galaxy. The $34\arcsec\times34\arcsec$ PC1 image of the galaxy is visible in the background.\label{fig:sectors}}
\end{figure}

This goal is achieved by fitting ``in parallel'' the MGE model of equation~(\ref{eq:surf_twist}), in polar coordinates, to a certain number $N_{\rm sec}$ of photometric profiles, measured along sectors uniformly spaced in angle from the major axis to the minor axis (sectors in the four quadrants are averaged together before fitting). All profiles include the central pixel and proper integration of the Gaussians on that pixel was performed for an accurate data-model comparison. The sampling of the photometry along the sectors is now logarithmically spaced in the \emph{elliptical} radius $m'^2=x'^2+y'^2/q'^2$, where $q'$ is a representative axial ratio of the galaxy isophotes (see Fig.~\ref{fig:sectors}). This latter detail ensures that the requirement stated in the previous paragraph is precisely satisfied also for flattened objects. This way of sampling galaxy photometry is a time proven practice common to most photometry packages (e.g., Jedrzejewski 1987). In the fit the same relative error is assigned to all measurements as in the 1D case, following equation~(\ref{eq:chi2new}).

The actual photometric measurements are obtained by selecting sets of pixels on the image, based on their angular position and logarithmic radius ranges. Once the pixels have been selected, the luminosity weighted average coordinates of each photometric point are evaluated from the pixels themselves and \emph{not} from the original sectors. This ensures that the coordinates of sectors at the border, or containing masked pixels, are properly computed and, even more importantly, it permits an accurate modeling of the galaxy surface brightness distribution close to the centre, where the discrete nature of pixels cannot be neglected. In particular \emph{no} surface brightness interpolation is performed in the centre, but the observed pixel values are used in the fit. To guarantee a proper sampling of the image, both in the nucleus and at large radii, 24 sectors have been adopted for every decade in radius \emph{in pixels units} (this corresponds to the usual factor 1.1 separation between successive radial intervals).

With these choices, when fitting a galaxy with slowly varying ellipticity with radius (as is often the case for single component objects like ellipticals), if the input guess $q'$ for the axial ratio is close to the galaxy average value, the procedure essentially reduces to $N_{\rm sec}$ 1D similar fits and converges very quickly by varying almost only the $\sigma_j$ as done in Section~\ref{sec:1d}. If the axial ratio is not constant or the isophotes are not ellipses (as in the case of spirals), both the $\sigma_j$ and $q'_j$ will have to be optimised together. In practice we still apply steps (i)-(iii) of Section~\ref{sec:1d}, with the only difference that we also fit the nonlinear parameters $q'_j$ in step (ii). To make the fit more robust, bound constraints are given on the variables: $R_{\rm min}<\sigma_j<R_{\rm max}$ and $0<q'_j<1$. Experience on a number of complex objects has shown that the above procedure will still converge very quickly and automatically to an excellent MGE fit to the photometry, since the initial guess for the solution is already good enough to permit the Levenberg-Marquardt optimisation algorithm to converge to the global minimum.

\begin{figure}
\fig{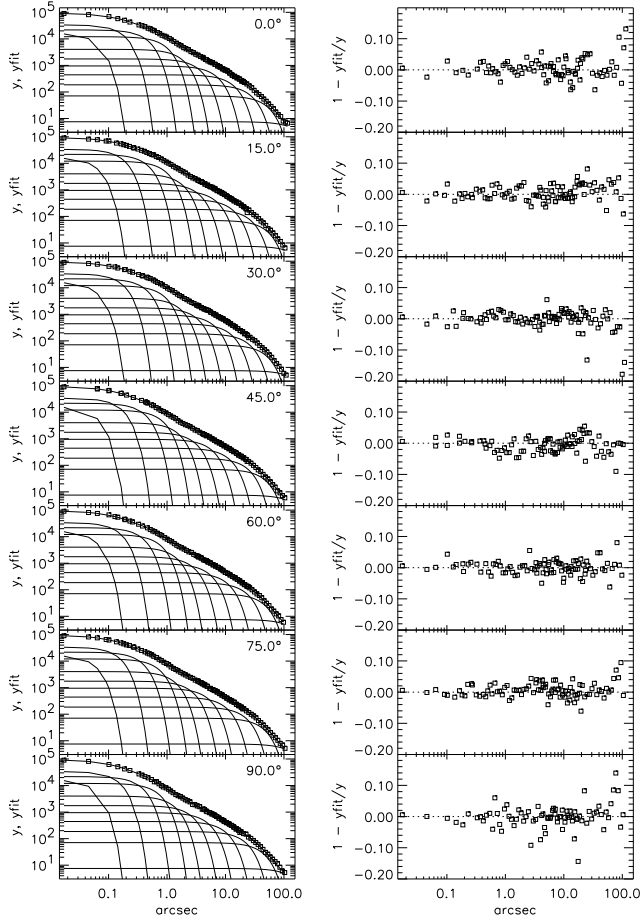}
\caption{\emph{Left Panels:} Comparison between the WFPC2/F814W + ground-based $I$-band photometry of M32 (open squares) and the corresponding $N=11$ Gaussians MGE best-fitting model (solid line). The individual convolved Gaussians components are also shown. From top to bottom the profiles are measured along $5^\circ$ wide sectors, linearly spaced in angle between the major ($0^\circ$) and the minor axis ($90^\circ$). Only 7 representative sectors of the of the $N_{\rm sec}=19$ sectors actually used in the fit are shown here. \emph{Right Panels:} radial variation of the relative error along the profiles. The RMS error is 2.2 per cent. Most of this scatter is due to the resolution of M32 into stars.
\label{fig:m32_profiles}}
\end{figure}

\begin{figure}
\fig{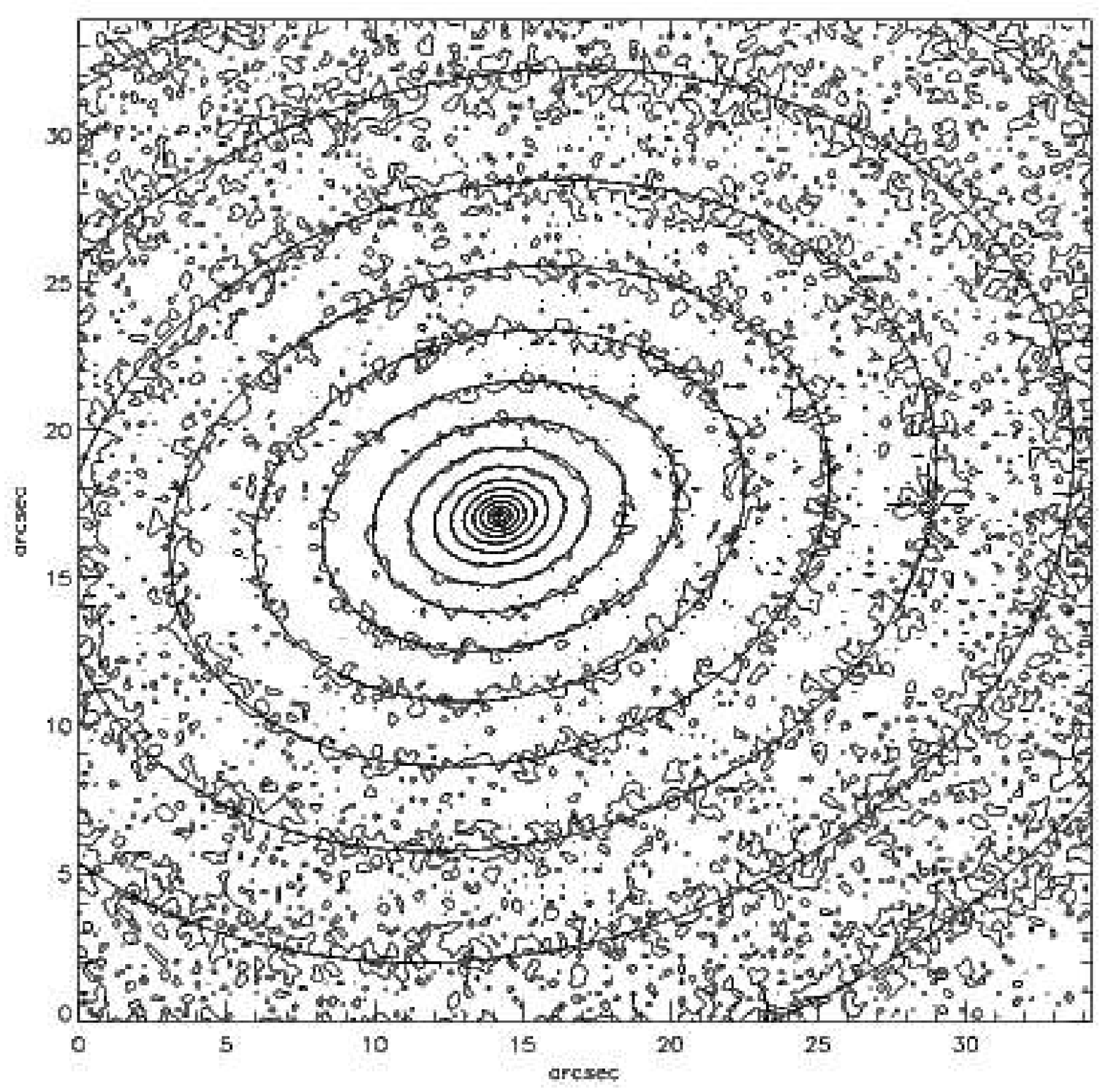}
\fig{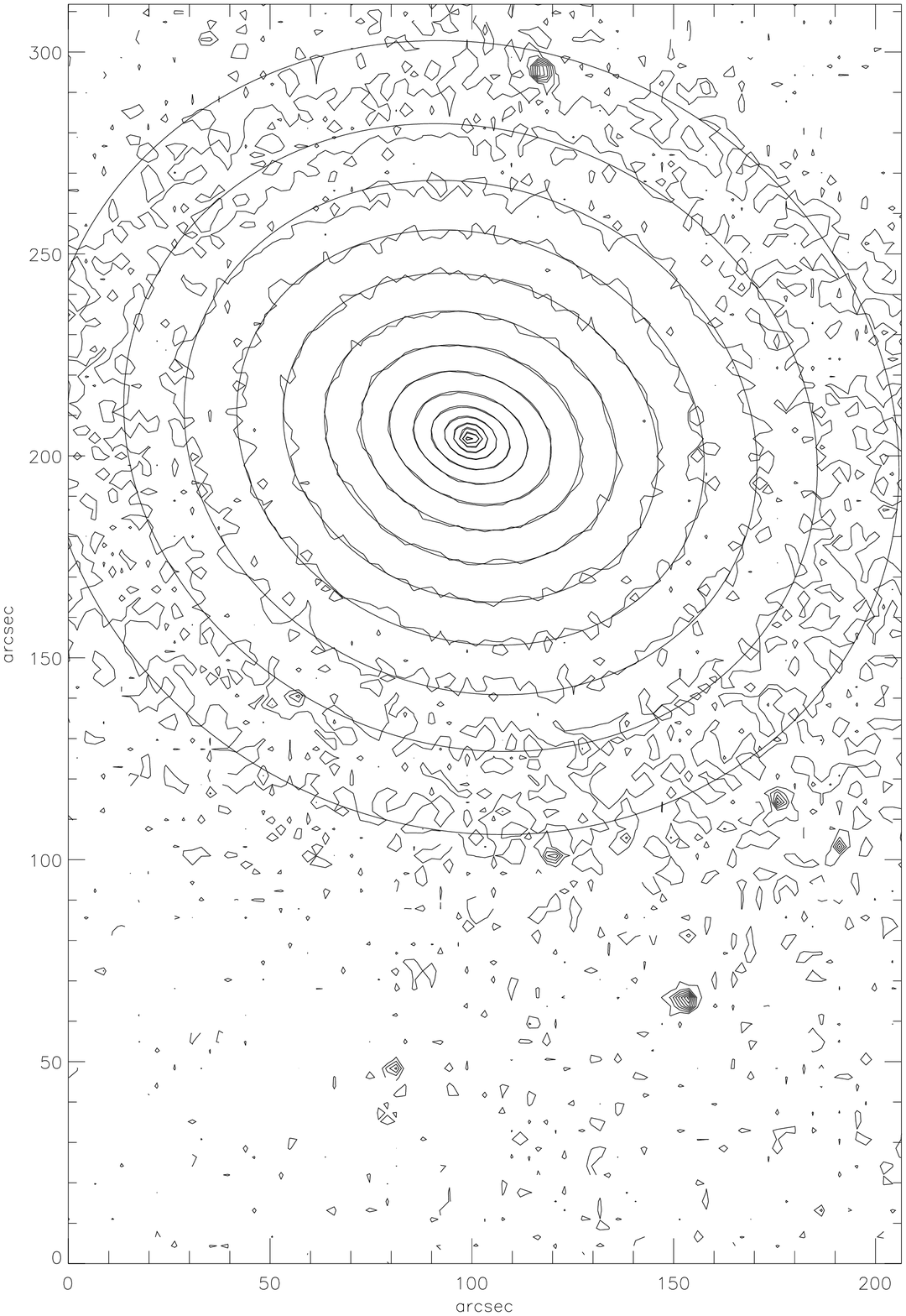}
\caption{\emph{Top panel:} Contour map of the central $34\arcsec\times34\arcsec$ of the (1280 s) WFPC2/F814W ($I$-band) image of M32. The noisy appearance of this high signal-to-noise HST image is due to the resolution of the galaxy into stars.
\emph{Bottom panel:} Same as in upper panel for the $200\arcsec\times300\arcsec$  ground-based (1 s) INT $I$-band image. Superimposed on the two plots are the  contours of the intrinsic MGE surface brightness, convolved with the proper PSF. In this plot, as in the following ones the contours are logarithmically spaced, but otherwise arbitrary.
\label{fig:m32_contours}}
\end{figure}

As an example we present in Figures~\ref{fig:m32_profiles} the profiles\footnote{A value $R_{\rm min}=[\sqrt{2}+\sinh^{-1}(1)]/6\simeq0.38$ pixels has been assigned in the plots to the radius of the central pixel, corresponding to the unweighted average radius measured from the centre of the pixel.} of an MGE fit to the elliptical galaxy M32 (cf. Bendinelli et al.\ 1992). The model was fit simultaneously to a WFPC2/F814W image and an $I$-band ground-based one (Peletier 1993). The $N=11$ Gaussians MGE fit for this galaxy was obtained with $N_{\rm sec}=19$ sectors, spaced by $5^\circ$, covering the whole galaxy, from the major to the minor axis. As expected the Gaussians succession is very regular and predictable, as in the 1D MGE fit previously shown in Fig.~\ref{fig:mge_fit_1d_m32}. The comparison between the contours of the model and the data is presented in Fig.~\ref{fig:m32_contours}. Similarly to what was done by Emsellem et al.\ (1994a), to determine the number of Gaussian components in the fit we start e.g., from $N=10$ Gaussians and keep adding Gaussians and restarting the fit, until the best-fitting $\chi^2$ stops decreasing appreciably (e.g., less than 1 per cent).

Instead of convolving the MGE model with the different PSFs during the fit we simply exclude the radial range that is affected by PSF effects, of the lower resolution image, and only convolve the model with the HST PSF. This is done to avoid spoiling the fit to the high resolution HST image by any PSF mismatch in the ground-based data. Sky subtraction and relative flux calibration of the frames is standard for the MGE method as for any other photometric modeling (e.g., Franx et al.\ 1989) and will not be discussed here in detail. The sky subtraction for the ground-based image was determined by requiring the outer profile to fall of as a power-law; the sky level and relative calibration of the HST image was adjusted to ensure agreement in the overlap region.

\begin{figure}
\fig{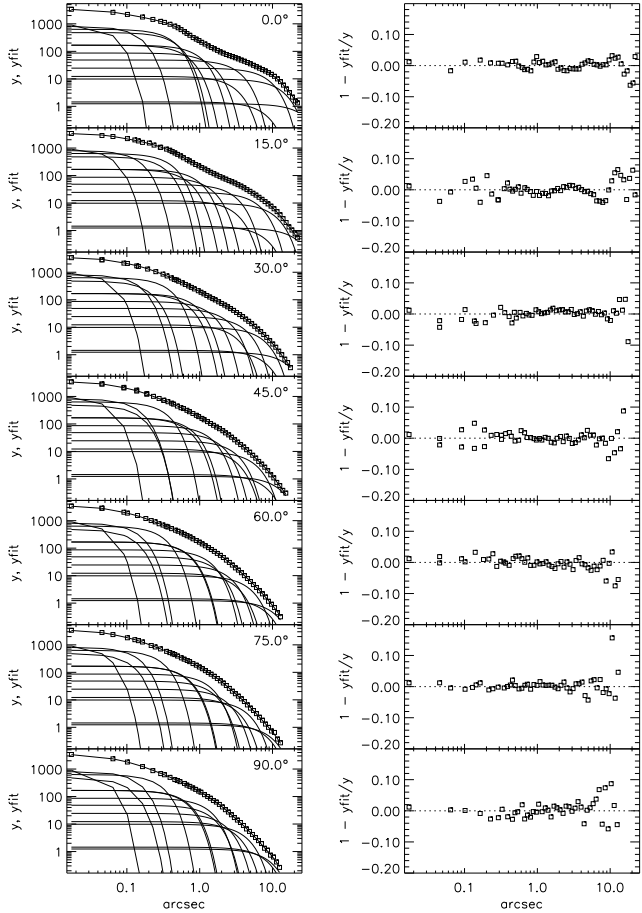}
\caption{Same as in Fig.~\ref{fig:m32_profiles} for the $N=13$ Gaussians MGE fit to the WFPC2/F814W photometry of NGC~4342. The RMS error is 1.8 per cent. Part of the data-model discrepancy at large radii along the major axis is due to a non perfect flatfielding of the PC1 CCD close to the border.
\label{fig:n4342_profiles}}
\end{figure}

\begin{figure}
\fig{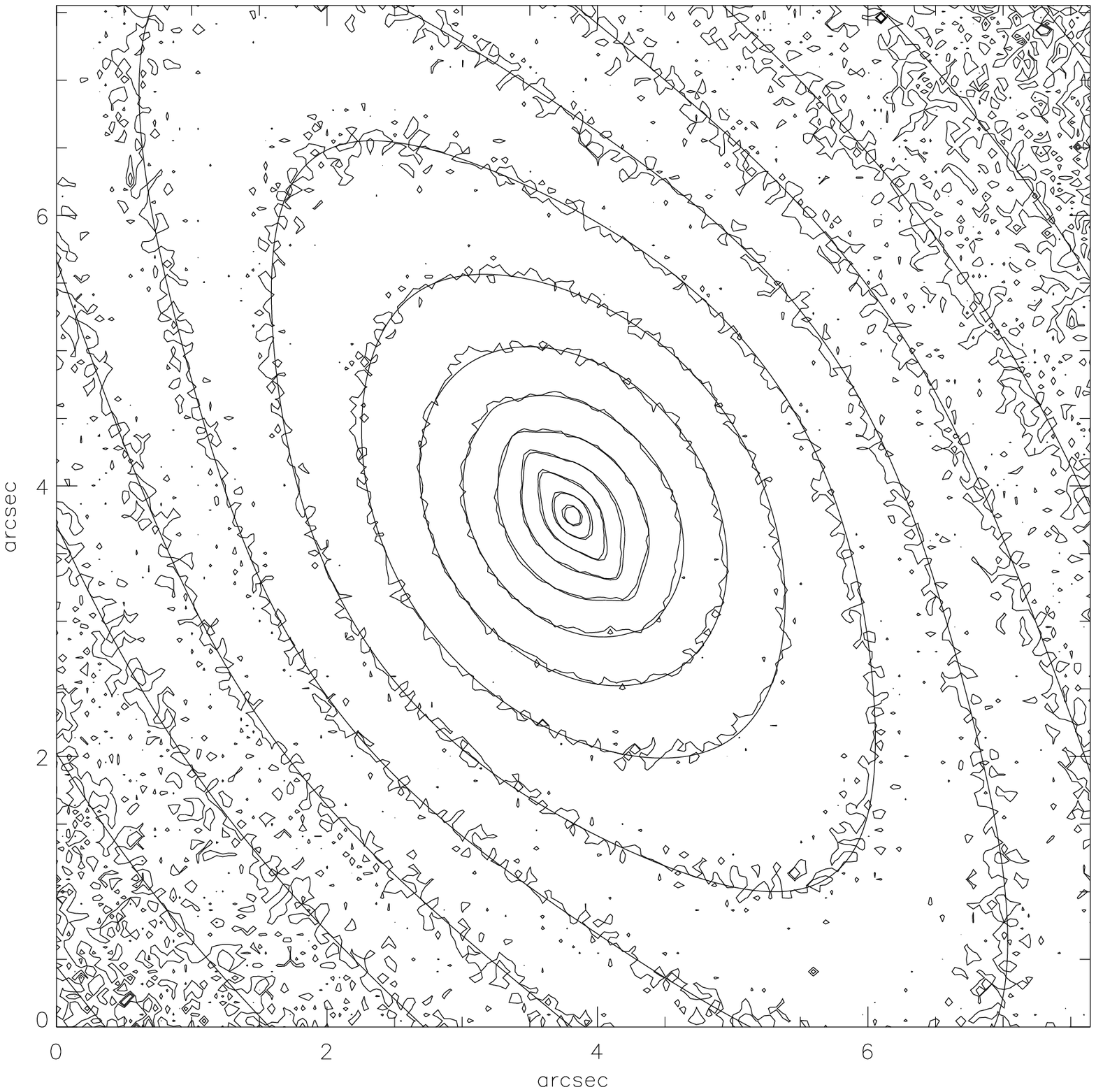}
\fig{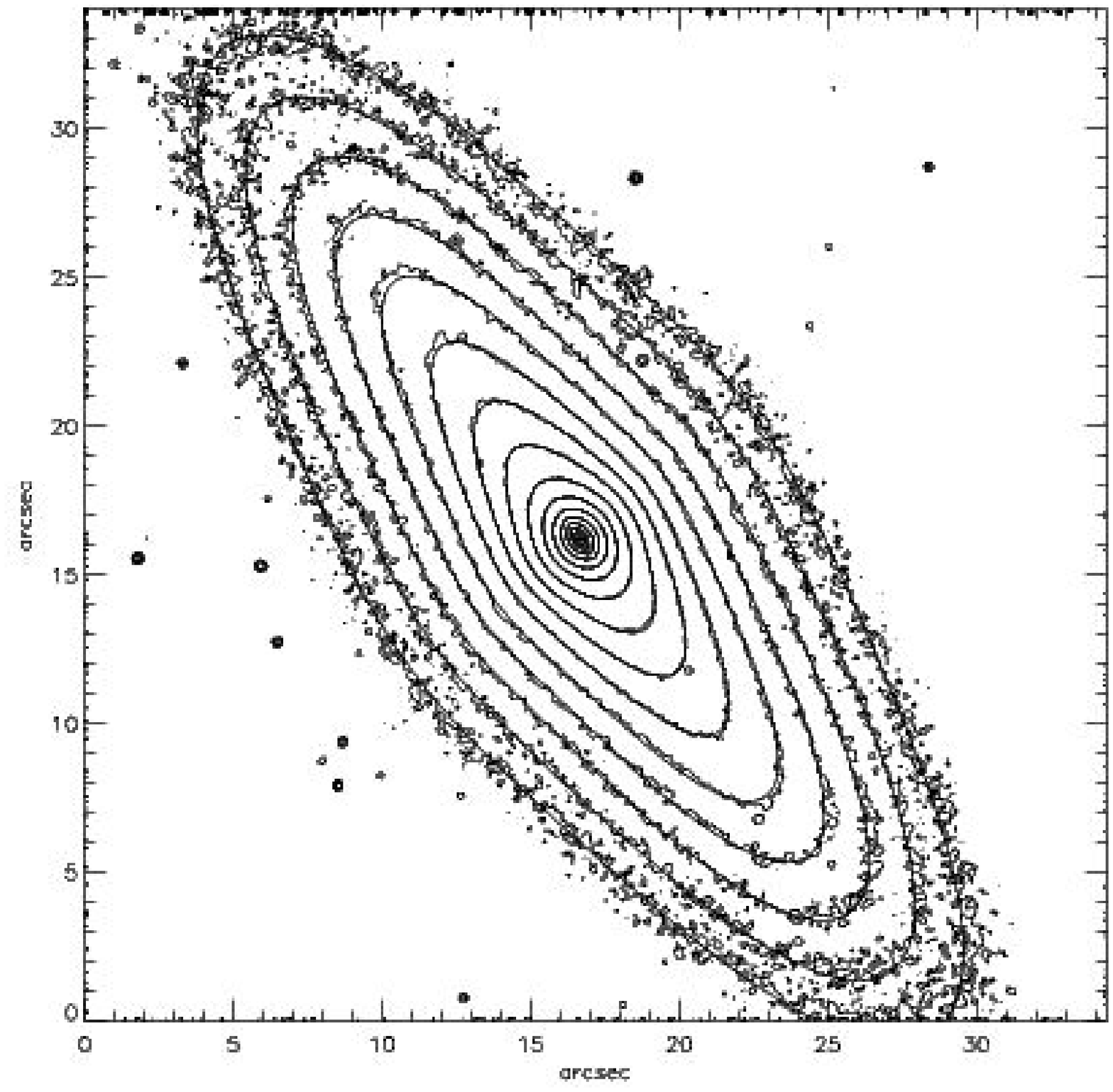}
\caption{Contour maps of the (200 s) WFPC2/F814W ($I$-band) image of NGC~4342 at two different scales: $8\arcsec\times8\arcsec$ (\emph{top panel}) and $34\arcsec\times34\arcsec$ (\emph{bottom panel}). Superimposed on the two plots are the  contours of the intrinsic MGE surface brightness (Fig.~\ref{fig:n4342_profiles}), convolved with the WFPC2 PSF. This figure can be compared with Fig.~1 in Cretton \& van den Bosch (1999).
\label{fig:n4342_contours}}
\end{figure}

An MGE fit to the more ``difficult'' S0 galaxy NGC~4342, which has a small nuclear stellar disk, in addition to a main outer disk (van den Bosch et al.\ 1998) is shown in Fig.~\ref{fig:n4342_profiles}. Also in this case the MGE fit with  $N=13$ Gaussians to the  WFPC2/F814W image of this galaxy was obtained with $N_{\rm sec}=19$ sectors, spaced by $5^\circ$, covering the whole galaxy, from the major to the minor axis. A comparison between the contours of our MGE model and of the observed image is shown in Fig.~\ref{fig:n4342_contours}.

\begin{figure}
\fig{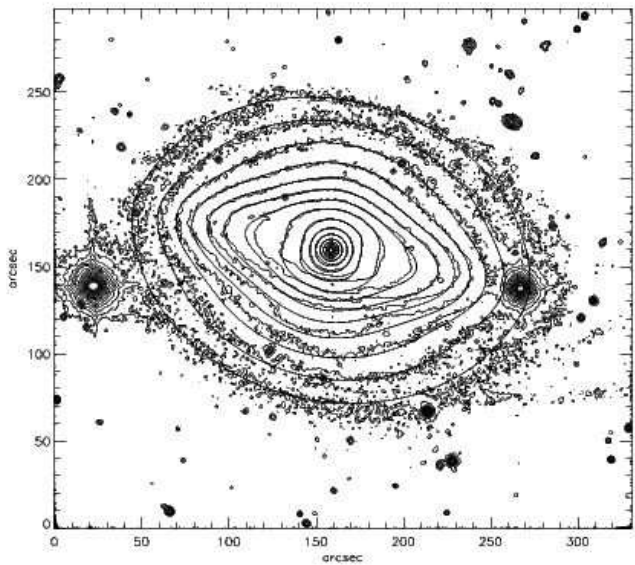}
\caption{Contour plot of a $V$-band ground-based image of the Sa galaxy NGC~4698, taken with the Vatican Advanced Technology Telescope (VATT). Overlaid are the contours of the best-fitting MGE model for this galaxy, obtained without imposing the usual positivity constraints on the Gaussians. Only in this way also the non-convex central isophotes can be reproduced. The prominent dust lanes in the bottom part of the image and the bright stars have been masked during the MGE fit.
\label{fig:ngc4698_mge}}
\end{figure}

In Fig.~\ref{fig:ngc4698_mge} is presented an example of an MGE fit to the Sa galaxy NGC~4698, which has non-convex isophotes (Bertola et al.\ 1999; Cappellari 2000) and for this reason necessarily cannot be reproduced by a sum of \emph{positive} Gaussians. In this case the positivity constraint for the $L_j$ has been eliminated and the linear part of the Gaussian fit has been performed using SVD as described in Section~\ref{sec:1d}.

NGC~4698 has a disk with prominent dust lanes. The techniques that can be used to determine an MGE model of a dusty galaxy are almost the same one can use to correct standard photometric measurements for dust effects (e.g., Carollo et al.\ 1997). While performing an MGE fit one can mask (exclude from the fit) the galaxy regions that are affected by clumpy dust absorptions. This method clearly works better if some symmetries can be assumed for the underlying galaxy. This was done to obtain the MGE model of Fig.~\ref{fig:ngc4698_mge}, with constant PA. MGE models like this can also be used to estimate and correct for dust absorption effects (Emsellem 1995). When images in more than one band are available, a colour excess map [e.g., $E(V-I)$] can be used to perform a first-order correction for dust absorption as done by Cappellari et al.\ (2002) to determine an MGE model of IC~1459.

The MGE fitting method discussed in this paper was already successfully applied to another sample of ten galaxies in Cappellari (2000). The method was also tested on twenty elliptical and lenticular galaxies with HST/WFPC2 photometry, taken from the SAURON representative sample (de Zeeuw et al.\ 2002). These models will be presented elsewhere. We have generally found that the RMS error of our MGE models is dominated by the noise or by intrinsic asymmetries in the photometry, and is of the order of $\sim1$ per cent or less.
In most cases a good fit is obtained in about one minute on the same 1 GHz machine that took many hours to obtain a poorer fit with the brute force approach.

We notice that the adopted MGE fit approach also has the advantage that it is very simple to fit together multiple photometric profiles obtained from different images. It is also easy to combine high resolution images with published photometric profiles at large radii. In fact during the fit the program only needs to know the measured values of the surface brightness and the corresponding polar coordinates, and does not have to deal with multiple images directly.

\subsection{A fully linear alternative 2D algorithm\label{sec:2d_lin}}

An alternative approach to the 2D MGE fit of an image, using Gaussians with the same PA and the same centre, consists of transforming the whole problem into a NNLS problem, that can be solved with the \emph{ad-hoc} Lawson \& Hanson (1974) algorithm. This can be done by generating a large grid of all possible Gaussians that may end up in the final solution. We sample $\sigma_j$ logarithmically with $R_{\rm min}<\sigma_j<R_{\rm max}$; for every choice of $\sigma_j$ we sample $q'_j$ linearly with $0<q'_j<1$. The non-negative superposition of the Gaussians that best fit the data is then found with NNLS. This approach is memory intensive since a copy of the data has to be kept in memory for every Gaussian in the grid: a $100\times100$ grid requires $10^4$ copies of the photometric data, yet it is already feasible if only the sectors are used.

The advantage of this method is that it is \emph{guaranteed} to find the lowest $\chi^2$ solution (although it may not be unique) within the accuracy imposed by the adopted grid of Gaussians. A big disadvantage is that in practice many more Gaussians are used by the NNLS algorithm to produce a fit with the same $\chi^2$ as found with our nonlinear algorithm of Section~\ref{sec:2d}. This is in general an important point, since one often needs to use the MGE parametrization to perform subsequent calculations whose speed depends on the number of Gaussians: the computation of the potential scales linearly with the number $N$ of Gaussians, while for example the solution of the Jeans equations scales as $N^2$ (Emsellem et al.\ 1994a).

As an example we have used a $100\times100$ grid of $(\sigma,q')$ values ($10^4$ Gaussians) to produce an MGE model for NGC~4342 using the same input photometry of Section~\ref{sec:2d}.
The NNLS algorithm gave a best solution with 38 positive Gaussians and an error only slightly smaller than that of the 13 Gaussians nonlinear solution of Fig.~\ref{fig:n4342_profiles} (RMS error of 1.7 per cent, to be compared with 1.8 per cent for the nonlinear solution).

Since the NNLS solution is ill-conditioned it may be possible to extract a subsample from the 38 Gaussians that gives essentially the same $\chi^2_{\rm best}$ of the best solution. In particular one would like to find the \emph{smallest} set of Gaussians that still verifies $\chi^2 < (1+\epsilon)\; \chi^2_{\rm best}$, where $\epsilon$ represents a chosen fractional tolerance. Unfortunately this is a combinatorial minimization problem whose exact solution would require (today) a very large amount of time even with 38 Gaussians. We have thus adopted the following approximate method:
\begin{enumerate}
    \item solve the NNLS problem obtained by fitting the whole $(\sigma,q')$ grid of Gaussians $G_j$ to the photometric data and define $\chi^2_{\rm best}=\chi^2$;

    \item construct a set $S$ containing the positive Gaussians $G_j$ from the previous solution, with $j=1,\ldots,P$;

    \item for $j=1,\ldots,P$ solve a new NNLS problem, excluding $G_j$ from $S$, and record the $\chi_j^2$ obtained in the solution. Define $\chi^2_{j,{\rm min}}=\min\{\chi^2_1,\ldots,\chi^2_P\}$;

    \item if $\chi^2_{j,{\rm min}} < (1+\epsilon)\; \chi^2_{\rm best}$, where $\epsilon$ is the maximum accepted fractional error in $\chi^2$, then eliminate $G_j$ from $S$, update the number $P\leftarrow P-1$, and go back to step (iii), otherwise return $S$.
\end{enumerate}

Running the above fully linear algorithm, we have been able to find (in a time 400$\times$ longer) a set of 14 Gaussians that produced a $\chi^2$, in the NGC~4342 fit, not larger than that of the 13 Gaussians nonlinear solution. In general we have found that this linear algorithm is not competitive, both in speed and number of Gaussians for a given $\chi^2$, with the one described in Section~\ref{sec:2d}. Its extreme robustness however is very attractive and it can be very useful to test that a given solution cannot be significantly improved, or in cases where the final number of Gaussians is not an issue.

The solution obtained by the fully linear MGE fitting method can be further improved by the nonlinear fitting method and the steps (ii)-(iv) outlined above can be iterated once more. We have implemented all this and found that it works very well, although the extra complication of combining the two methods is not required most of the times.

\subsection{MGE fits of galaxies with isophote twist}

In Sections~\ref{sec:2d}--\ref{sec:2d_lin} we have shown how to fit MGE models to galaxies having isophotes with four-fold symmetry (constant PA with radius). Here we briefly discuss how to produce an MGE fit to galaxies whose isophotes twist with radius.

The extension of the nonlinear fitting method of Section~\ref{sec:2d} to galaxies with isophote PA that varies with radius comes very natural from the previous discussion. It has been shown that, in a single component object like an elliptical, every Gaussian contributes to the surface brightness essentially only at an elliptical radius $m'$ close to its $\sigma$. Consider for simplicity the case of constant axial ratio $q'$. As before the $\sigma_j$ will have to be logarithmically spaced and all the $q'_j$ can be set equal to a representative galaxy axial ratio. A very good starting guess for PA $\psi_j$ of the Gaussian $G_j$ will then be given by the PA of the isophote with semimajor axis $a=\sigma_j$. The PA at the various radii can be measured before fitting with standard photometry packages. As before for the constant-PA fitting problem, the starting guess with constant axial ratio provides a good approximation to the actual model fit with varying isophote ellipticity and the nonlinear optimisation method converges to the global minimum.

\begin{figure}
\fig{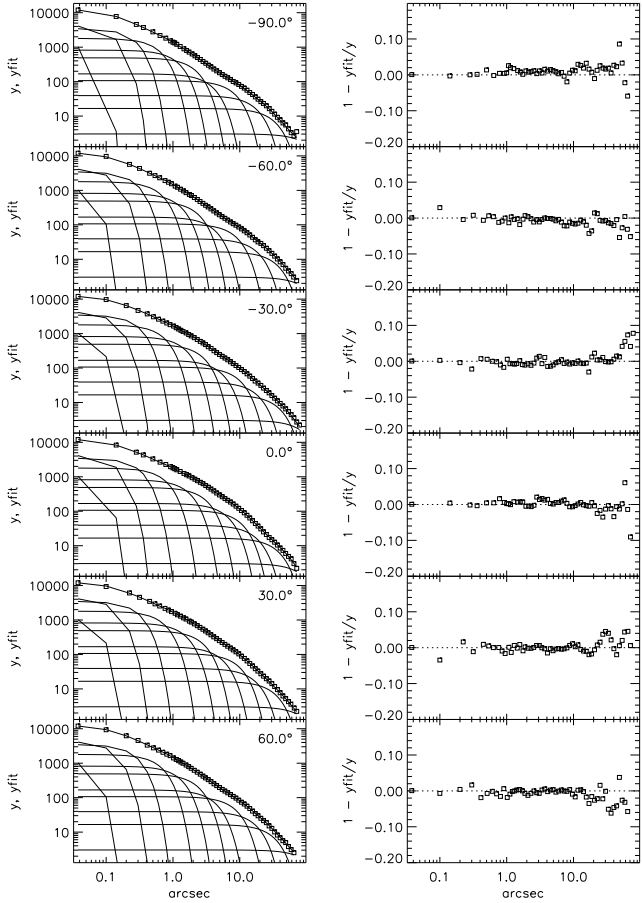}
\caption{Same as in Fig.~\ref{fig:m32_profiles} for the $N=11$ Gaussians MGE fit to the WFPC2/F702W photometry of NGC~5831. Only 6 equally spaced representative profiles are shown, of the $N_{\rm sec}=36$ sectors actually used in the fit. $0^\circ$ corresponds to the PA of the central isophotes. The RMS error is 1.2 per cent.}
\label{fig:n5831_profiles}
\end{figure}

\begin{figure}
\fig{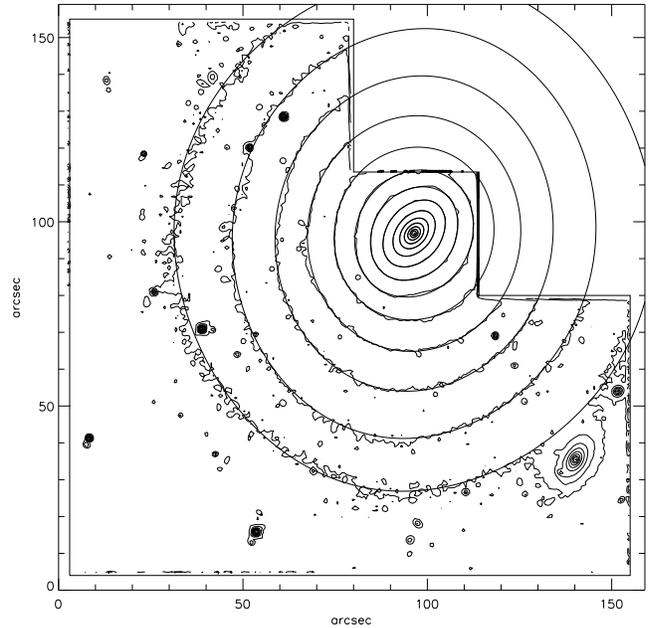}
\caption{Contour map for the 160$''$$\times$160$''$ (1000 s) WFPC2/F702W image of elliptical galaxy NGC~5831. Note the $\sim35^\circ$ isophote twist of this galaxy. Overlaid are the contours of the MGE best-fitting model whose profiles are shown in Fig.~\ref{fig:n5831_profiles}.}
\label{fig:n5831_contours}
\end{figure}

Fig.~\ref{fig:n5831_profiles} presents the profiles of the $N=11$ Gaussians MGE best-fitting model for a WFPC2/F702W image of the elliptical (E3) galaxy NGC~5831, that presents a $\sim35^\circ$ isophote twist in the range $0-40''$ (Peletier et al.\ 1990; Binney \& Merrifield 1998, figure 4.37). The profiles have been measured along 36 sectors $5^\circ$ wide, covering the whole galaxy image from $-90^\circ$ to $90^\circ$. Since we are fitting a centre-symmetric model, the counts measured within sectors symmetric with respect to the centre have been averaged before fitting. A very good MGE model can be obtained for this galaxy. As expected, although the PA is different for the various fitted components, the Gaussians succession is still regular and predictable as in the case of constant PA of Fig.~\ref{fig:m32_profiles}. A comparison between the contours of the data and of the convolved MGE model is shown in Fig.~\ref{fig:n5831_contours}.

\begin{figure}
\fig{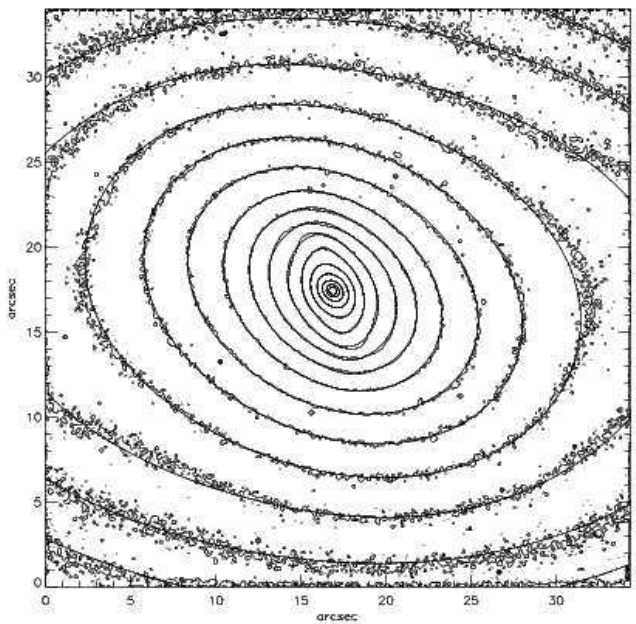}
\caption{Contour map of the central 34$''$$\times$34$''$ of the (730 s) WFPC2/F814W ($I$-band) image of the barred S0 galaxy NGC~2950. The best-fitting MGE model is overlaid. Note the sharply twisted and non elliptical isophotes due to a double bar within this galaxy. $N=7$ Gaussians have been used in the MGE model and the resulting RMS error is 2.1 per cent.}
\label{fig:n2950}
\end{figure}

Fig.~\ref{fig:n2950} shows an MGE fit to the central region of the S0 barred galaxy NGC~2950. This object presents a sharp $\sim60^\circ$ isophote twist inside the central $20''$, due to the presence of a double bar component (Friedli et al.\ 1996; Rest et al.\ 2001). An $N=7$ Gaussians MGE model is able to reproduce very well the central surface brightness of this object, including the strongly twisted and non elliptical isophotes, with the exception of the slightly ``spiral-like'' isophote shape that peaks at an elliptical radius of $\sim7''$ and is clearly visible from the contour plot.

The fully-linear MGE fitting method of Section~\ref{sec:2d_lin} can also be easily extended to deal with isophote twisting. It is simply necessary to add Gaussians with a range of PA $\psi$, linearly spaced from  $-90^\circ$ to $90^\circ$, in the representative library of Gaussian, for every $(\sigma,q')$ combination. The NNLS fit to the photometry will be performed with a three-dimensional grid $(\sigma,q',\psi)$ of Gaussians. The only problem of this approach is that the memory requirements will increase considerably, but a 30$\times$30$\times$30 grid of Gaussians can already be solved with current averages PCs.

\section{Availability}

The complete IDL\footnote{http://www.rsinc.com} source code implementing the linear and nonlinear MGE fitting algorithms, described in this paper, both with and without isophote twist, together with manual and usage examples, can be found at http://www.strw.leidenuniv.nl/$\sim$mcappell/idl.

\section{Conclusions}

The MGE method is one of the few simple parametrizations that are general enough to reproduce the surface brightness of realistic multicomponent galaxies. In addition many dynamical and photometric quantities can be evaluated easily and accurately when the density is expressed in MGE form.

In this paper we have described a simple yet powerful algorithm that reduces the process of generating an MGE fit to multiple galaxy images to a simple, fast and automatic task. We have provided examples of its practical use and have tested it by accurately reproducing the photometry of a relatively large sample of galaxies, both with and without isophote twists. We have also compared some of our fits with previously obtained photometric models.

We have also described an alternative algorithm that, although currently less practical, due to the larger computing power requirements, is guaranteed to converge to the minimum $\chi^2$ solution within the accuracy imposed by an adopted grid of parameters in the solution space.

These algorithms have been implemented in an IDL program that can produce an MGE model starting from the observed images of a galaxy, requiring the user to only input the coordinates of the galaxy centre, the PA and a characteristic flattening. Multiple resolution images (e.g., ground-based and HST) can easily be fitted together, in one single step. The complete IDL source code implementing the algorithms described in this paper is made publicly available.

\section*{Acknowledgements}

It is a pleasure to thank Eric Emsellem for stimulating discussions that provided the motivation to undertake this project. I wish to thank Tim de Zeeuw and Ellen Verolme for careful reading of the manuscript and very useful comments and suggestions. I thank the anonymous referee for constructive comments.
I am grateful to Jos\'e Funes and Alessandro Pizzella for obtaining the CCD image shown in Fig.~\ref{fig:ngc4698_mge} and to Claudia Scarlata for providing the reduced version of it. That image was obtained with the VATT: the Alice P.\ Lennon Telescope and the Thomas J.\ Bannan Astrophysics Facility.

\label{lastpage}


\begin{thebibliography}{}

\bibitem[\protect\citeauthoryear{Bendinelli}{1991}]{ben91} Bendinelli O., 1991, ApJ, 366, 599

\bibitem[\protect\citeauthoryear{Bendinelli et al.}{1992}]{ben92} Bendinelli O.,
Zavatti F., Parmeggiani G., Djorgovski S., 1992, AJ,  103, 110

\bibitem[\protect\citeauthoryear{Bertin \& Arnouts}{1996}]{ber96} Bertin E., Arnouts S., 1996, A\&AS, 117, 393

\bibitem[\protect\citeauthoryear{Bertola et al.}{1991}]{ber91} Bertola F., Bettoni D.,
Danziger J., Sadler E., Sparke L., de Zeeuw P.T., 1991, ApJ,  373, 369

\bibitem[\protect\citeauthoryear{Bertola et al.}{1999}]{ber99} Bertola F., Corsini E.M., Beltr{\' a}n J.C.V., Pizzella A., Sarzi M., Cappellari M., Funes S.J., 1999, ApJ, 519, L127

\bibitem[\protect\citeauthoryear{Binney}{1985}]{bin85} Binney J., 1985, MNRAS,  212, 767

\bibitem[\protect\citeauthoryear{Binney \& Merrifield}{1998}]{bin98} Binney J., Merrifield M., 1998, Galactic Astronomy (Princeton: Princeton Univ. Press)

\bibitem[\protect\citeauthoryear{Binney \& Tremaine}{1987}]{bin87} Binney J., Tremaine S., 1987, Galactic Dynamics (Princeton: Princeton Univ. Press)

\bibitem[\protect\citeauthoryear{Cappellari}{2000}]{cap00} Cappellari M., 2000, PhD thesis, Padova University

\bibitem[\protect\citeauthoryear{Cappellari et al.}{2002}]{cap02} Cappellari M., Verolme E.K., van der Marel R.P., Verdoes Kleijn G.A., Illingworth G.D., Franx M., Carollo C.M., de Zeeuw P.T., 2002, MNRAS, submitted

\bibitem[\protect\citeauthoryear{Carollo et al.}{1997}]{car97} Carollo C.M., Franx M., Illingworth G.D., Forbes D.A., 1997, ApJ, 481, 710

\bibitem[\protect\citeauthoryear{Carter}{1978}]{car78} Carter D., 1978, MNRAS,  182, 797

\bibitem[\protect\citeauthoryear{Chandrasekhar}{1969}]{cha69} Chandrasekhar, S.\ 1969, Ellipsoidal Figures of Equilibrium, (New Haven: Yale University Press)

\bibitem[\protect\citeauthoryear{Contopoulos}{1956}]{con56} Contopoulos G., 1956, Zeitschrift f\"ur Astrophysik,  39, 126

\bibitem[\protect\citeauthoryear{Cretton \& van den Bosch}{1999}]{cre99a} Cretton N., van den Bosch F.C., 1999, ApJ, 514, 704

\bibitem[\protect\citeauthoryear{de Zeeuw \& Franx}{1989}]{dzw89} de Zeeuw P.T., Franx M., 1989, ApJ,  343, 617

\bibitem[\protect\citeauthoryear{de Zeeuw et at.}{2002}]{dzw02} de Zeeuw P.T. et al., 2002, MNRAS, 329, 513

\bibitem[\protect\citeauthoryear{Emsellem}{1995}]{ems95} Emsellem E., 1995, A\&A, 303, 673

\bibitem[\protect\citeauthoryear{Emsellem, Monnet \& Bacon}{1994}]{ems94a} Emsellem E., Monnet G., Bacon R.,  1994a, A\&A, 285, 723

\bibitem[\protect\citeauthoryear{Emsellem et al.}{1994}]{ems94b} Emsellem E., Monnet G.,  Bacon R.,   Nieto J.-L.,  1994b, A\&A, 285, 739

\bibitem[\protect\citeauthoryear{Emsellem et al.}{1996}]{ems96} Emsellem E., Bacon R., Monnet G.,  Poulain P.,  1996, A\&A, 312, 777

\bibitem[\protect\citeauthoryear{Emsellem, Dejonghe \& Bacon}{1999}]{ems99} Emsellem E., Dejonghe H.,  Bacon R.,  1999, MNRAS, 303, 495

\bibitem[\protect\citeauthoryear{Franx}{1988}]{fra88} Franx M., 1988, MNRAS,  231, 285

\bibitem[\protect\citeauthoryear{Franx et al.}{1989}]{fra89} Franx M., Illingworth G.,
Heckman T., 1989, AJ,  98, 538

\bibitem[\protect\citeauthoryear{Friedli et al.}{1996}]{fri96} Friedli D., Wozniak H.,
Rieke M., Martinet L., Bratschi P., 1996, A\&AS,  118, 461

\bibitem[\protect\citeauthoryear{Hunter \& Qian}{1993}]{hun93} Hunter C., Qian E. 1993, MNRAS, 202, 401

\bibitem[\protect\citeauthoryear{Jedrzejewski}{1987}]{jed87} Jedrzejewski R.I., 1987, MNRAS, 226, 747

\bibitem[\protect\citeauthoryear{Kent}{1984}]{ken84} Kent S.M., 1984, ApJS,  56, 105

\bibitem[\protect\citeauthoryear{Lauer}{1985}]{lau85} Lauer T.R., 1985, ApJS,  57,
473

\bibitem[\protect\citeauthoryear{Lauer et al.}{1995}]{lau95a} Lauer, T.R. et al., 1995, AJ, 110, 2622

\bibitem[\protect\citeauthoryear{Lawson \& Hanson}{1974}]{law74} Lawson C.L., Hanson R., 1974, Solving Least Squares Problems; 1995, 2nd ed. (Englewood Cliffs, NJ: Prentice-Hall)

\bibitem[\protect\citeauthoryear{Monnet, Bacon \& Emsellem}{1992}]{mon92} Monnet G., Bacon R., Emsellem E., 1992, A\&A, 253, 366

\bibitem[\protect\citeauthoryear{Mor\'e, Garbow \& Hillstrom}{1980}]{mor80} Mor\'e J.J., Garbow B.S., Hillstrom K.E., 1980, User Guide for MINPACK-1, Argonne National Laboratory Report ANL-80-74

\bibitem[\protect\citeauthoryear{Peletier}{1993}]{pel93} Peletier R.F., 1993, A\&A, 271, 51

\bibitem[\protect\citeauthoryear{Peletier et al.}{1990}]{pel90} Peletier R.F., Davies
R.L., Illingworth G.D., Davis L.E., Cawson M., 1990, AJ,  100, 1091

\bibitem[\protect\citeauthoryear{Press et al.}{1992}]{pre92} Press W.H., Teukolsky S.A., Vetterling W.T., Flannery B.P., 1992, Numerical Recipes in FORTRAN 77 (2d ed; Cambridge: Cambridge Univ. Press)

\bibitem[\protect\citeauthoryear{Rest et al.}{2001}]{res01} Rest A., van den Bosch
F.C., Jaffe W., Tran H., Tsvetanov Z., Ford H.C., Davies J., Schafer J.,
2001, AJ,  121, 2431

\bibitem[\protect\citeauthoryear{Rybicki}{1986}]{ryb86} Rybicki G.B., 1986, in Structure and Dynamics of Elliptical Galaxies, ed. P.T.\ de Zeeuw (Dordrecht: Reidel), 397

\bibitem[\protect\citeauthoryear{Stark}{1977}]{sta77} Stark A.A., 1977, ApJ,  213, 368

\bibitem[\protect\citeauthoryear{van den Bosch \& Emsellem}{1998}]{vdbem98} van den Bosch F.C., Emsellem E., 1998, MNRAS, 298, 267

\bibitem[\protect\citeauthoryear{van den Bosch, Jaffe \& van der Marel}{1998}]{vdb98} van den Bosch F.C., Jaffe W., van der Marel R.P. 1998, MNRAS, 293, 343

\bibitem[\protect\citeauthoryear{van der Marel \& Franx}{1993}]{vdm93} van der Marel, R.P., \& Franx, M. 1993, ApJ, 407, 525

\bibitem[\protect\citeauthoryear{van der Marel et al.}{1998}]{vdm98a}  van der Marel R.P., Cretton N., de Zeeuw P.T., Rix H.-W, 1998, ApJ, 493, 613

\bibitem[\protect\citeauthoryear{Williams}{1981}]{wil81} Williams T.B., 1981, ApJ, 244, 458

\end{thebibliography}
\end{document}